\documentclass[12pt,tightenlines,nofootinbib]{revtex4-2}
\usepackage{amsfonts,amsmath,amssymb}
\usepackage{txfonts}
\usepackage{graphicx,color}
\usepackage{wrapfig,lipsum}
\usepackage{floatrow,float}
\usepackage{epstopdf}
\usepackage{hyperref}
\usepackage[normalem]{ulem}
\usepackage{enumitem}
\usepackage{mathrsfs,mathtools}

\DeclareMathOperator{\e}{erf}

\begin{document}
	
	\title{Grid Cell Percolation}
	\author{Yuri Dabaghian}
	\affiliation{Department of Neurology, The University of Texas McGovern Medical School, 6431 Fannin St, Houston, TX 77030\\
		$^{*}$e-mail: Yuri.A.Dabaghian@uth.tmc.edu}
	\vspace{17 mm}
	\date{\today}
	
\begin{abstract}
	Grid cells play a principal role in enabling mammalian cognitive representations of ambient environments. 
	The key property of these cells---the regular arrangement of their firing fields---is commonly viewed as 
	means for establishing spatial scales or encoding specific locations. However, using grid cells' spiking
	outputs for deducing spatial orderliness proves to be a strenuous task, due to fairly irregular activation
	patterns triggered by the animal's sporadic visits to the grid fields. The following discussion addresses 
	statistical mechanisms enabling emergent regularity of grid cell firing activity, from the perspective of
	percolation theory. In particular, it is shown that the range of neurophysiological parameters required 
	for spiking percolation phenomena matches experimental data, which points at biological viability of the 
	percolation approach and casts a new light on the role of grid cells in organizing the hippocampal map. 
\end{abstract}

\keywords{grid cells $|$ percolation $|$ learning and memory $|$ hippocampo-cortical network} 	
\maketitle
\newpage

\section{Introduction and motivation}
\label{sec:int}

\begin{figure}
	\centering
	\includegraphics[scale=0.8]{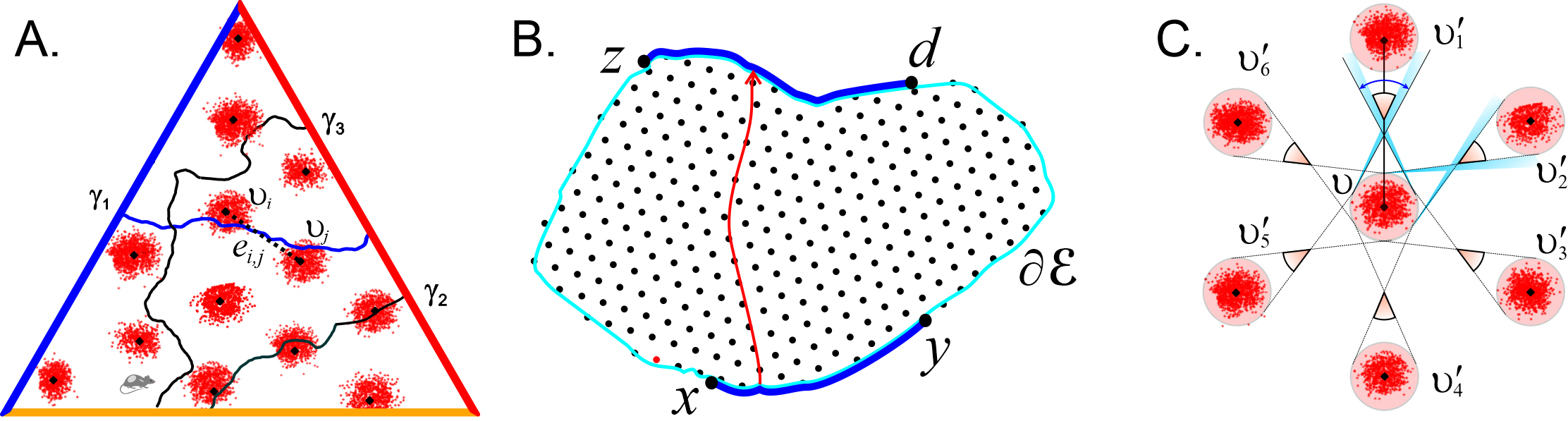}
	\caption{{\footnotesize
			\textbf{Grid cells}. (\textbf{A}). Grid fields form a hexagonal lattice embedded into a $3\times3$
			m triangular enclosure. Vertexes of the lattice are shown by black dots. The diameter of each field
			is comparable to the animal's body size. Three curves, $\gamma_1$, $\gamma_2$ and $\gamma_3$, 
			represent paths segments extending from one side of a triangular environment to another. The path
			$\gamma_1$ crosses the grid fields $\upsilon_i$ and $\upsilon_j$ eliciting spikes in both, thus 
			opening the vertexes $v_i$ and $v_j$ and instantiating the edge $e_{ij}$ between them (dashed line).
			Path $\gamma_2$ percolates through the grid fields in sequence, without omissions, $\gamma_3$ avoids
			grid fields.
			\textbf{B}. A percolation theory setup: a path extending from one side of a lattice domain 
			$V_{\mathcal{E}}$, a boundary segment ($[a,b]\in\partial\mathcal{E}$) to another ($[z,d]\in\partial
			\mathcal{E}$) may open vertexes or edges with fixed probabilities $p_v$ and $p_e$.
			\textbf{C}. The range of directions that lead from a grid field $\upsilon$ to its six immediate
			neighbors, $\upsilon_1,\ldots,\upsilon_6$, are marked by pink. Directions along which a straight
			line escapes between the fields are marked by blue. For $\xi_g=1/2$, the escape directions 
			disappear.
	}}
	\label{fig:gc}
\end{figure}

Cognitive representation of space is sustained by the spiking activity of ``spatially tuned'' neurons,
such as hippocampal place cells, head direction cells, parietal cells, border cells, and others \cite{Kropff,Grieves}. 
A particularly curious pattern of activity is exhibited by the grid cells in the rats' Medial Entorhinal Cortex
(MEC) that fire in compact domains centered at the vertexes of a triangular lattice, tiling the navigated
environment \cite{Hafting} (Fig.~\ref{fig:gc}A). 
The exact principles by which these cells contribute to spatial awareness remain a matter of debate. It is
commonly assumed that MEC outputs are used to represent the animal's ongoing location and to establish global
spatial metrics \cite{Bush2,MoserM}. However, extracting these structures from the spike train patterns is a
complex task: since the animal can visit one firing field at a time, the sequences of grid cell responses 
depend on the shape of the rat's trajectory and can be highly intermittent. In absence of simple universal
decoding algorithms, the effect produced by the grid cells in the downstream networks may depend primarily on
activation frequency: persistently firing cells contribute most, while the ones that activate sporadically 
produce smaller impacts \cite{Syntax}. The maximal frequency of a given grid cell's responses is achieved over
periods when the animal runs through its firing fields in sequence, without omissions (Fig.~\ref{fig:gc}A). 
Question is, under which conditions such contiguous firing can be produced in a given environment, how common
are the regularly firing cells and so forth. 

Curiously, these questions are reminiscent of the problems addressed in percolation theory, which describes 
propagation of diffusive substances (liquids or gasses) through porous media. The key question addressed by
the theory is whether a permeable domain $\mathcal{E}$ allows diffusive leaks from one side of its boundary
to another \cite{Grimmet}, i.e., whether the trickling through the pores can form uninterrupted sequences 
connecting the opposite sides, or \textit{percolate}\footnote{Throughout the text, terminological definitions
	are given in \textit{italics}.} through, $\mathcal{E}$. 

Of particular interest for the following discussion are mathematical models of percolation, in which physical
media is represented by a segment of a regular lattice $V$ enclosed within a domain $\mathcal{E}$. Depending
on the setup, either the vertexes $v$ or the edges $e$ of the lattice $V_{\mathcal{E}}$ represent the leaking
pores, which may ``open'' or ``close'' with some fixed probabilities \cite{Grimmet,Kesten} (Fig.~\ref{fig:gc}B).
A key result of the theory is that exceeding critical thresholds (for triangular lattices, $p_v^{\ast}=0.5$ for
the vertexes and $p_e^{\ast}=2\sin(\pi/18)\approx0.35$ for the edges), marks the onset of a \textit{percolating
	phase}, in which uninterrupted sequences of open sites, connecting opposite sides of the lattice 
$V_{\mathcal{E}}$, become statistically common \cite{Wierman}.

The analogy with the grid cells can be formalized as follows. Consider a triangular lattice $V_{\mathcal{E}}^g$
with vertexes centered at the firing fields of a cell $g$. A vertex $v_i^g$ opens if the cell $g$ fires at
the corresponding field $\upsilon_i^g=\upsilon(v_i^g)$. If the rat runs consecutively through two neighboring 
fields, e.g, from $\upsilon_i^g$ to $\upsilon_j^g$ on Fig.~\ref{fig:gc}A, eliciting spikes in both, then the 
edge $e_{i,j}^g$ between them also opens. If a path $\gamma$ induces a sequence of conjoint open edges,
\begin{equation}
	\mathsf{G}_g(\gamma)=\{e_{i_1,i_2}^g,e_{i_2,i_3}^g,\ldots,e_{v_{i-1},i_k}^g\},
	\label{G0}
\end{equation} 
(and hence runs through a series of open sites $v_{i_1}^g,v_{i_2}^g,\ldots,v_{i_k}^g$), it will be said to 
percolate $g$. The spiking pattern of a cell $g$ triggered by the rat's moves can then be described by a
sequence of open vertexes and edges, i.e., represented by discretized the path (\ref{G0}).

In the following, it is shown that, for a certain scope of firing parameters, the paths $\gamma$ extending
through $\mathcal{E}$ may systematically percolate groups of grid cells, which may then play a particular role
in representing spatial information. The activity of MEC network can therefore be studied from the perspective
of identifying such high-impact cells, understanding their role in representing the navigated paths, testing 
whether the parameters required for percolation are physiologically viable and so forth.

\section{Results}
\label{sec:prelim}
\subsection*{Preliminary estimates} Grid cell percolation depends on the probability with which a generic
trajectory runs into the grid fields and the probability of eliciting spiking responses. The former is 
controlled by the ratio between the field size $D_g$ and the grid spacing $a_g$,
\begin{equation}
	\xi_g=D_g/a_g,
	\label{xi}
\end{equation}
while the latter depends on the maximal firing rate $A_g$ and the animal's speed $s$.

\textit{Lattice parameter} $\xi_g$ defines the range of directions that lead from a grid field to one of
its immediate neighbors. If the gap between fields is wider than the field size, $\xi_g<1/2$, then a finite
fraction of straight directions originating at a given grid field form ``escape corridors''---passageways 
in-between the surrounding fields (Fig.~\ref{fig:gc}C). At $\xi_g= 1/2$ such directions disappear, suggesting
that $\xi_g\gtrsim 1/2$ is required for enforcing the percolation. 
However, this requirement must be strengthened further, for two reasons. First, the trajectory cannot not just
brush on the field's side, where the firing rate is too small---it should pass sufficiently close to the center,
to induce reliable spiking responses. Second, the ``empirical'' size of a field is defined by the lengths of the
typical paths that run though it, rather than the field's diameter.
A simple correction to (\ref{xi}) can hence be obtained by replacing the diameter $D_g$ with the length of an
average chord cutting through the field, $\bar{l}_g=\pi D_g/4$ \cite{Kellerer1,Coleman1}, which yields $\xi_g
\gtrsim 2/\pi$ (Fig.~\ref{fig:gc}B). 

If $\xi_g$ grows further, the angular domains leading from field to field begin to overlap. To maintain 
unambiguity of representation, the lattice parameter (\ref{xi}) should remain close the marginal value,
\begin{equation}
	\xi_g^{\ast}\approx 2/\pi,
	\label{isom}
\end{equation}
which matches the experimentally observed ``isometric relation'' \cite{Hafting,Neher}.

\textit{The opening probabilities} on a grid field lattice depend on the parameters of neuronal activity and
the animal's moves. If the maximal spiking rate of a cell $g$ is $A_g$, then the mean rate is
\begin{equation}
	\bar{\lambda}_g=C \frac{A_g D_g}{\bar{s}},
	\label{lambdabar}
\end{equation}
where $C\approx0.06$ is a geometric coefficient, and $\bar{s}$ is the mean traversal speed (see Appendix,
Sec.~\ref{sec:met}). Experimentally, the grid field sizes sampled along the ventro-dorsal axis of MEC range,
in smaller environments, from about $10$ cm to about $20$ cm \cite{Hafting,Stensola}, while the mean rates 
co-vary between $A_g\approx 21$ Hz to $A_g\approx 11$ Hz, i.e., the product $A_gD_g$ tends to increase as the
field sizes grow. In larger environments, the firing rates remain approximately same, while field sizes cover
the range $50 \lesssim D_g\lesssim 120$ cm, driving $\bar{\lambda}_g$ to higher values \cite{BrunG}. Thus, if
the rat spends $2-3$ seconds within a field, the expected firing probability,
\begin{equation*}
	\bar{p}_v^g=1-e^{-\bar{\lambda}_g},
\end{equation*}
is high---the corresponding cell fires almost certainly. In particular, the mean vertex opening probability
exceeds the critical value $p_v^{\ast}$, which suggests that vertex percolation may indeed take place in the
parahippocampal network. On the other hand, the expected probability of opening an edge $e$ can be estimated
from
\begin{equation*}
	\bar{p}_e^g=\bar{q}_e^g(\bar{p}_{v}^g)^2,
\end{equation*}
where $\bar{q}_e^g$ is the mean probability of reaching a grid field starting from its closest neighbor.
Geometrically, $\bar{q}_e^g$ depends primarily on the lattice parameter $\xi_g$ (Sec.~\ref{sec:met}).
For the experimentally observed $\xi_g\approx2/3$, the estimated value is about $\bar{q}_e^g(2/3)\approx 0.6$,
which, for high enough $\bar{p}_v^g$, reaffirms the possibility that some grid cells may be systematically
percolated during navigation.

\subsection*{Grid field percolation} These hypotheses can be tested by simulating rat's navigation in triangular
environments, for a set of spiking parameters. Experimentally, grid spacings $a_g$ range along the ventro-dorsal
axis of MEC from $0.3$ m to $1.2$ m in smaller environments \cite{Stensola,Hafting}, and from $1.7$ m to $3$ m 
and higher in larger environments \cite{BrunG}. The place field sizes grow accordingly, and since the effect of 
speed is stronger for smaller fields, percolation is least likely to occur in smaller lattices. 

\begin{figure}[h]
	\centering
	\includegraphics[scale=0.8]{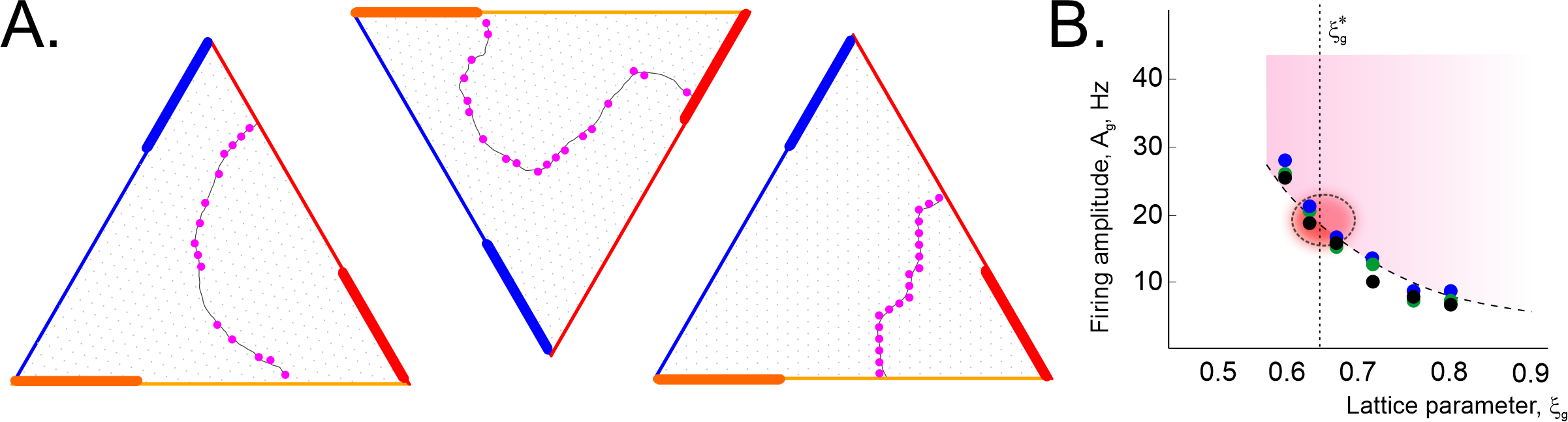}
	\caption{{\footnotesize\textbf{Grid field percolation}. 
			\textbf{A}. A segment of simulated trajectory (gray line) passing through a large $20\times20$ m
			triangular environment. Active vertexes are marked by pink circles. The first two paths are 
			non-percolating, the third path percolates the enclosure. Side bars indicate $L/3$ length scale.
			\textbf{B}. For a given set of parameters $(\xi_g,A_g)$, the onset of the percolation phase was
			scored when at least $90\%$ of grid cells were percolated by at least $5\%$ of cross-environment
			path segments, and $90\%$ of paths extending over at least $L/3$ percolate a grid cell. The latter
			condition defined the size of the simulated grid cell ensemble---about $N_g\gtrsim 200$ cells. 
			Dots of different colors mark values obtained using different exploratory trajectories
			that cover the environment evenly, without artificial favoring one part of the environment over
			the other. The dashed line separates two phases of the grid cell network's activity: the pairs of
			$(\xi_g,A_g)$-values on its right (pink area) induce percolation, while the values on its left do 
			not. Encircled area marks the domain of smallest $A_g$ that permit percolation at $\xi_g\approx
			\xi_g^{\ast}$. 
	}}
	\label{fig:perc1}
\end{figure}

For conservative estimates, the grid cell spacing in the simulations was therefore fixed at a lower value, $a_g
=60$ cm, while the lattice parameter varied from $\xi_g=0.8$ (fields almost abut) to $\xi_g=1/3$. The spatial
phases and orientations of the grid field lattices $V_{\mathcal{E}}^g$ were randomized to represent different 
possibilities for cells sampled along the ventro-dorsal axis of MEC. To maintain realistic dynamics of spiking
activity, the trajectories were generated by re-shaping experimentally recorded paths in open arenas, preserving
the observed speed of the animal. The length of each trajectory allowed producing at least $100$ path segments 
extending from one side of the environment to another.

The results show that paths crossing an equilateral triangular enclosure with side $L=6$ m start percolating 
grid cells as the lattice parameter and the firing amplitude exceed, respectively, $\xi_g\approx 0.6$ and $A_g
\gtrsim20-25$ Hz, independently from the lattice's shift and planar orientations (Fig.~\ref{fig:perc1}A,B). As
$\xi_g$ grows further, percolating paths start appearing at lower firing rates and quickly proliferate, densely
covering the navigated area. On the other hand, increasing $A_g$ allows inducing percolation at lower $\xi_g$.
The pairs $(\xi_g,A_g)$ for which the percolation becomes possible form a boundary that separates ``percolation
phase'' from the phase in which percolation is statistically suppressed (Fig.~\ref{fig:perc1}B). 

\begin{figure}[h]
	\centering
	\includegraphics[scale=0.8]{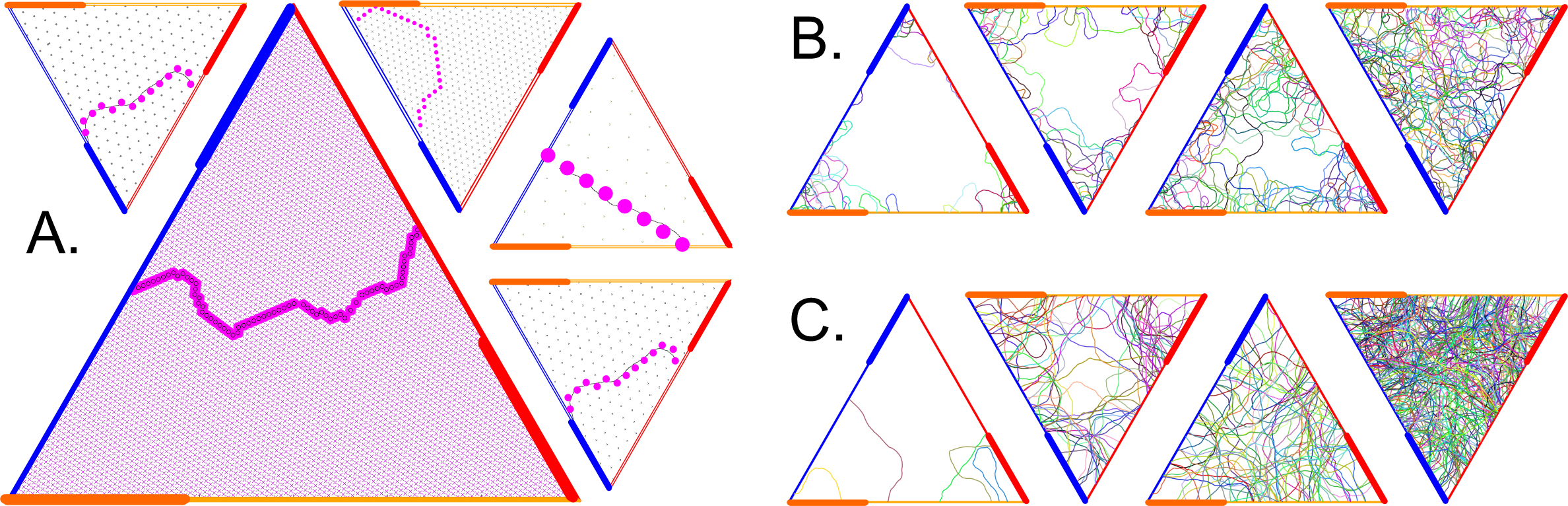}
	\caption{{\footnotesize\textbf{Scaling}. 
			\textbf{A}. As the lattice parameter exceeds $\xi_g\approx 2/3$, percolating paths begin to appear
			at all scales. Shown are triangular environments with side lengths $L=6$ m, $L=12$ m, $L=20$ m and
			$L=60$ m along with a percolating path examples.
			\textbf{B}. As the rate $A_g$ increases from $20$ Hz to $200$ Hz ($L=12$ m, $\xi_g\approx2/3$), 
			larger parts of the environment become percolated. 
			\textbf{C}. Increasing $\xi_g$ from $2/3$ to $0.8$ ($L=20$ m, $A_g\approx25$ Hz) boosts the 
			percolation. Shown are path segments percolating the same cell, each shown with its	own color. 
			Side bars indicate $L/3$ length scale.
	}}
	\label{fig:scl}
\end{figure}

Furthermore, once emerged, percolation becomes manifested at large scales, e.g., in triangular domains that 
differ in size by an order of magnitude (Fig.~\ref{fig:scl}A). The effect is strengthened as the cell's firing
amplitude $A_g$ grows, in a manner suggestive of a second-order phase transition controlled by two order 
parameters, $\xi_g$ and $A_g$ \cite{Watanabe,Stokes}.

Note however, that since higher firing rates are energetically costly, electrophysiologically recorded
magnitudes $A_g$ should remain close to minimal values that permit a suitable percolation level at a given
spatial scale. As shown on Fig.~\ref{fig:perc1}B, such estimated values also appear to match the experimental
data, which indicates physiological viability of the percolation model.

\subsection*{Spike lattice in the cognitive map} The results discussed above were obtained by modeling the 
rat's moves through the firing fields in the observed environment---an approach that helps visualizing the grid
cells' spiking patterns, but may not directly capture the organization of the underlying network computations
\cite{PLoS,Rev}. Understanding the latter requires placing the grid cells' activity into the context
of the brain's own representation of the environment---the cognitive map, encoded, \textit{inter alia}, by the
place cell and the head direction cell networks \cite{Kropff,Grieves}. The computational units enabling this
representation are the functionally interconnected groups of hippocampal place cells, $c_i$ 
\cite{Harris2,DrgBuz},
\begin{subequations}
	\label{cqs}
	\begin{align}
		\begin{split}
			\sigma_i&=[c_{i_0},c_{i_1},\ldots, c_{i_n}],
			\label{sigma}
		\end{split}
		\tag{5$\sigma$}
		\intertext{and head direction cells, $h_i$ \cite{Peyrache2,BrandonTh},}
		\begin{split}
			\eta_j&=[h_{j_1},h_{j_2},\ldots,h_{j_n}].
			\label{eta}
		\end{split}
		\tag{5$\eta$}
	\end{align}
\end{subequations}
As their constituent place and head direction cells, the assemblies (\ref{cqs}) are spatially selective and
highlight, respectively, basic locations $\upsilon_{\sigma_i}$ and angular domains $\upsilon_{\eta_j}$ 
\cite{DrgBuz,Harris2,Peyrache2,BrandonTh}. The relative arrangement of these ``fields'' defines the order in
which the assemblies ignite; knowing the latter allows decoding the animal's positions during active behavior
\cite{Jensen,Frank,Guger} and during the ``off-line'' memory explorations \cite{Karl,Johnson,Dragoi,Pfeiffer1}.
One can hence model hippocampal representation of the grid cells' firing patterns by the same principle: each 
individual grid field $\upsilon_{i}^g$ is encoded by those place cell assemblies, $\sigma_{i_1},\sigma_{i_2},
\ldots,\sigma_{i_k}$, whose fields are contained in $\upsilon_{i}^g$, i.e., by the place cells that exhibit 
coactivity with a given cell $g$ and each other. 

Computationally, the assemblies (\ref{cqs}) are commonly modeled as the cliques of a graph that represents 
recurrent functional connectivity in the network, e.g., of the \textit{cognitive graph} that represents the 
collaterals in the CA3 region of the hippocampus \cite{Muller,Burgess}.  Simulations show that such assemblies
form agglomerates, $\hat{\sigma}_i=\{\sigma_{i_1},\sigma_{i_2},\ldots,\sigma_{i_k}\}$, whose joint firing 
domains, $\upsilon_{\hat{\sigma}_i}$, cover the individual grid field regions. The corresponding combinations
of place- and grid cells can hence be viewed as the units encoding the \textit{spiking vertexes},
\begin{equation}
	\hat{\varv}_{i}^g=[\hat{\sigma}_{i},g],
	\label{v}
\end{equation} 
in the hippocampal cognitive map.  

The hexagonal order on these vertexes is then established by concomitant activity of head direction assemblies from
six ``preferred'' groups,
\begin{equation*}
	\mathfrak{h}^g=\{\hat{\eta}^g_1,\hat{\eta}^g_2,\ldots,\hat{\eta}^g_6\},
\end{equation*} 
that activate on the runs between pairs of neighboring grid fields, e.g., assemblies from $\hat{\eta}^g_1$ may
activate when the rat runs approximately from left to right, assemblies from $\hat{\eta}^g_2$ then become active
on the runs oriented $60^{\circ}$ from the left-right direction and so forth \cite{Peyrache2}. Correspondingly, 
the activity of $\eta$-assemblies from a particular $\hat{\eta}$-group that leads from a vertex $\hat{\varv}_i$
to a neighboring vertex $\hat{\varv}_j$ defines an \textit{spiking edge},
\begin{equation}
	\epsilon_{i,j|k}^{g}=\{\hat{\sigma}_i,\hat{\sigma}_j|\hat{\eta}_{k},g\}.
	\label{e}
\end{equation}
Together, the vertexes (\ref{v}) and the edges (\ref{e}) define segments of a \textit{spiking lattice} 
$\mathcal{V}_g$ embedded into the cognitive map. In the following, the superscript $g$ and subscript $k$ will
be used to distinguish contributions from different grid and head direction cells, and suppressed otherwise.

\subsection*{Percolation of spiking lattice} The ``intrinsic'' definition of the lattice elements (\ref{v}) and
(\ref{e}) given above leads to a natural generalization of the grid field percolation model. As the animal's 
trajectory $\gamma$ traverses a discrete sequence of $\sigma$-fields,
\begin{equation*}
	\Upsilon=\{\upsilon_{\sigma_1},\upsilon_{\sigma_2},\ldots\upsilon_{\sigma_n},\ldots\},
\end{equation*}
a ``firing trace'' of ignited place cell assemblies,
\begin{subequations}
	\label{trace}
	\begin{align}
		\begin{split}
			\digamma_{\sigma}(\gamma)\equiv(\sigma_1,\sigma_2,\ldots),
			\label{ptrace}
		\end{split}
		\tag{7$\sigma$}
		\intertext{is induced in the hippocampal network, along with a sequence of ignited head direction
			assemblies,}
		\begin{split}
			\digamma_{\eta}(\gamma)\equiv(\eta_1,\eta_2,\ldots).
			\label{htrace}
		\end{split}
		\tag{7$\eta$}
	\end{align}
\end{subequations}
The representation (\ref{ptrace}) of the navigated path \cite{Jensen,Frank,Guger} then allows defining 
\textit{spiking percolation} as follows: 
\vspace{7pt}
\begin{enumerate}[nosep]
	\item[\textbf{P1}.] A spiking vertex $\hat{\varv}_i$ \textit{opens} when its constituent cells activate,
	i.e., when a place cell assembly from its ``hippocampal base'' $\hat{\sigma}_{i}$ co-activates with the grid
	cell $g$;
	\item[\textbf{P2}.] Two consecutively opening, neighboring vertexes $\hat{\varv}_1$ and $\hat{\varv}_2$
	produce	an \textit{open spiking edge} in $\mathcal{V}_\mathcal{E}$ if the head direction assemblies from 
	a fixed	group $\hat{\eta}_{\ast}$ remain active on the run from $\hat{\sigma}_1$ to $\hat{\sigma}_2$;
	\item[\textbf{P3}.] The trace $\digamma_{\sigma}(\gamma)$ percolates through $\mathcal{V}_\mathcal{E}$ if
	it runs through a sequence of consecutively opening vertexes $\hat{\varv}_i$ without omissions.
\end{enumerate}
\vspace{7pt}
The grid field percolation discussed in $\mathsection2$ can be viewed as a geometric, ``pictorial'' 
representation of the spiking percolation if the observed animal moves and the pattern of grid fields are 
physiologically actualized, i.e., if place cells' activity marks every location of the rat and if the head
direction activity chaperones every move between neighboring grid fields\footnote{To simplify modeling, 
	movement direction was used as a proxy for the head direction, although physiologically these parameters
	not identical \cite{Raudies,Aff}}.

Simulations show that the required output is provided by as few as $N_c\gtrsim 100$ active place cells per unit
area ($1m\times 1m$; experimentally observed numbers are higher by an order of magnitude \cite{Syntax}), with
typical firing parameters (mean place field size $D_c\approx 24$ cm, mean firing rate amplitudes $A_c\approx20$
Hz). Furthermore, $N_h\gtrsim 60$ head direction cells firing with the amplitude $A_h\approx 20$ Hz over $D_h=
20^{\circ}$ fields form lattice direction groups $\hat{\eta}_i$ (about $10$ cells each) that distinguish runs 
of the simulated rat between different pairs of neighboring grid fields, which demonstrates that spiking
percolation can occur within physiological range of parameters.
The resulting series of conjoint open spike edges 
\begin{equation*}
	\mathfrak{G}_g(\gamma)=\{\epsilon_{i_1,i_2}^g,\epsilon_{i_2,i_3}^g,\ldots,\epsilon_{v_{i-1},i_k}^g\},
\end{equation*}
form an intrinsic, spike-lattice representation of the grid field lattice path (\ref{G0}) at the scale defined
by the lattice constant $a_g$.

\subsection*{Path integration} A number of models were built to explain the role of the grid cells in the 
animal's capacity to optimize navigation using a cognitive map of ambient environment \cite{Sav1,Val1,McNPth}.
The mechanisms by which parahippocampal and entorhinal networks learn to represent space and retrieve the 
obtained information through autonomous network dynamics remain debated \cite{Samsonovich,Val1,McNaughton3}.
A model suggested in \cite{Hasselmo1,Hasselmo2} implements the required hippocampal replays using persistently
firing head direction cells that drive grid cells' firing from vertex to vertex, which, in turn, activate the 
corresponding place cell assemblies in spatial order. If the network is trained according to the coactivity
between different types of neurons along the navigated path, e.g., $\delta W_{\eta,\sigma}\propto\sum_i\vec{p}_
{\eta_i}\vec{p}_{\sigma_i},$ where $\vec{p}_{\eta}$ and $\vec{p}_{\sigma}$ are the population activity vectors,
then the learned patterns can be reproduced autonomously in the retrieval phase.
For example, head direction firing can be induced by the place cells that start spiking at a position $\sigma$,
\begin{equation}
	\vec{p}_{\eta}=\sum_{\sigma} W_{\eta,\sigma}\vec{p}_{\sigma},
	\label{p}
\end{equation}
which can then drive the grid cell membrane oscillations, thus generating hippocampal activity at the net step
and so forth \cite{Hasselmo1,Hasselmo2}.

\begin{figure}[h]
	\centering
	\includegraphics[scale=0.84]{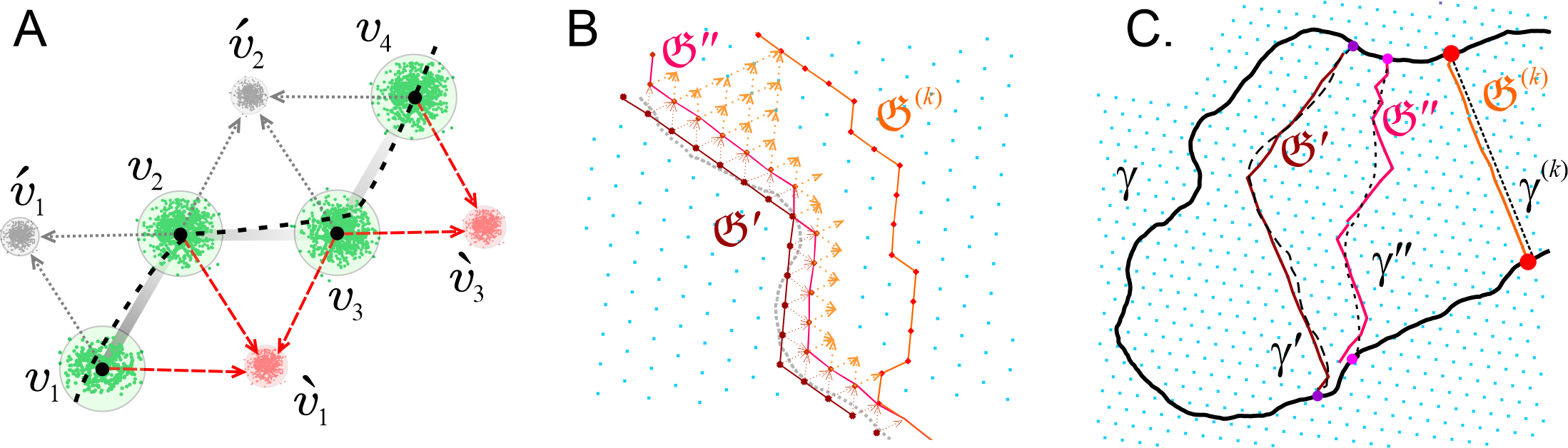}
	\caption{{\footnotesize\textbf{Path deformation}.
			\textbf{A}. The activity can transition from an open edge $\epsilon_{1,2}$ connecting two vertexes 
			$\varv_1$ and $\varv_2$ to its side vertex $\grave{\varv}_{1}$ (red field), thus opening the edges
			$\epsilon_{1,\grave{1}}$ and $\epsilon_{2,\grave{1}}$. Next, the activity can propagate from the 
			adjacent open edge, $\epsilon_{2,3}$, to the side vertex $\grave{\varv}_{2}=\grave{\varv}_{1}$, 
			opening the edge $\epsilon_{3,\grave{1}}$, and so forth. The induced shifts are driven towards one
			side of $\mathfrak{G}$, to allow continuous attractor dynamics. Dashed curve represents a segment
			of the rat's trajectory.
			\textbf{B}. A series of transformations (\ref{shift}) can be used to deform the percolating path
			$\mathfrak{G}_0$ over the lattice, $\mathfrak{G}\to\to\ldots$. 
			\textbf{C}. Consecutive deformations of discretized paths ($\mathfrak{G}'$, $\mathfrak{G}''$, etc.)
			can be used to propagate replays of alternative trajectories ($\gamma'$, $\gamma''$, etc.) along 
			the lattice and to produce lattice geodesics, i.e., shortest paths between lattice vertexes, e.g.
			$\gamma^{(k)}$.
	}}
	\label{fig:pp}
\end{figure}

From the perspective of this discussion, the network should be trained on the percolating paths only, which can
then be reproduced in replays. Aslo note that consecutive activation of two hippocampal assemblies $\sigma$ and
$\sigma'$ induced by a persistently firing head direction group $\eta$ may be viewed geometrically as a 
transition of activity between two adjacent $\sigma$-locations aligned along the $\eta$-direction \cite{Aff}. 
The Hasselmo model \cite{Hasselmo1,Hasselmo2} is hence based on using persistent head direction firing to guide
place cell activity from a grid vertex to a neighboring one. As it turns out, this mechanism can be generalized
to implement the transitions not only along the learned lattice edges, but also to probe their vicinities, which
significantly extends the scope of the model.

Consider an edge $\epsilon_{i,i+1|k}$ linking two open vertexes, $\varv_{i}$ and $\varv_{i+1}$, along a spike
lattice direction $\hat{\eta}_k$. Let $T_{\hat{\eta}_l\hat{\eta}_k}$ be the matrix permuting the assemblies
from two lattice direction groups, $\hat{\eta}_k$ and $\hat{\eta}_l$ \cite{Means}. Then the adjusted weight 
matrix
\begin{equation}
	T_{\hat{\eta}_l\hat{\eta}_k}(\sigma)W_{\hat{\eta}_k,\sigma}=W_{\hat{\eta}_l,\sigma},
	\label{T}
\end{equation} 
applied at the location $\sigma$ in (\ref{p}), redirects the persistent head direction activity from 
$\hat{\eta}_i$ to $\hat{\eta}_j$ (physiologically, this operation may be interpreted as a cortical or thalamic
switch \cite{Rikhye}). Two transformations of the weight matrices (\ref{T}) applied at the ends of an open edge
$\epsilon_{i,i+1|k}$, 
\begin{equation}
	T_{\hat{\eta}_{l}\hat{\eta}_k}(\sigma_i)W_{\hat{\eta}_k,\sigma_i}+T_{\hat{\eta}_{l'}\hat{\eta}_k}(\sigma_{i+1})
	W_{\hat{\eta}_k,\sigma_{i+1}}=W_{\eta,\grave{\sigma}_i},
	\label{shift}
\end{equation}
yield the weight matrix that funnels the activity from $\varv_{i}$ and $\varv_{i+1}$ to a side vertex, $\grave
{\varv}_{i}$ (Fig.~\ref{fig:pp}A). Same mechanism can then reroute the activity from the next open edge, 
$\epsilon_{i+1,i+2}$, to its side vertex $\grave{\varv}_{i+1}$, and so forth. 

From the percolation model's perspective, activation of the side vertexes also opens the edges that lead to 
these vertexes, which geometrically amounts to ``indenting'' the percolated lattice paths (Fig.~\ref{fig:pp}).
A series of such indentations can deform and shift the representation of the original path over the spike 
lattice, $\mathfrak{G}_g\to\mathfrak{G}_g^{\prime}\to\ldots\to\mathfrak{G}_g^{(k)}\to\ldots$, i.e., induce 
geometrically deformed lattice paths that can generate hippocampal replays of alternative, ``virtual'' 
trajectories and thus guide spatial exploration (Fig.~\ref{fig:pp}B, \cite{Sanders}). 

In particular, the possibility of deforming generic percolating paths allows establishing \textit{lattice
geodesics}---shortest chains of edges connecting pairs of vertexes. Hippocampal replay of the shortest path 
between the underlying spatial locations $\sigma$ and $\sigma'$ may account, e.g., for the animal's ability 
to run from its current position straight to the nest, which is a key manifestation of path integration 
\cite{Maaswinkel}. Another implication is that the shortest paths across the spiking lattice define a global
spatial metric---the discrete-geodesic distances between pairs of locations \cite{MoserM}, (Fig.~\ref{fig:pp}C).

Note that the transformations (\ref{shift}) can be used to redirect the activity to both sides of the open edge
series (Fig.~\ref{fig:pp}A). However, if the head direction cells' firing is to form a single ``activity bump''
defining a compact range of angles \cite{Bassett,StringerHD}, then the activity should be driven to one side of
the percolated path $\mathfrak{G}_g$ only. Gradual shifts of the activity bump in the head direction network 
along a deformed path $\mathfrak{G}_g$ are then consistent with continuous reorientations of the animal's head.

\section{Discussion}
\label{sec:disc}

Grid cell activity is commonly studied from the perspective of extracting position codes and spatial metrics 
from the combinatorics of \textit{ad hoc} defined grid field indexes \cite{Bush2}. Hereby, most models assume,
tacitly or explicitly, that a generic grid cell readily conveys spatial regularity of the grid field layouts 
to downstream networks through spiking outputs, over each navigated path. However, direct simulations show that,
over a given traveled route $\gamma$, most grid cells exhibit irregular spiking patterns that reflect the 
sequence in which their firing fields were visited, rather than the abstracted order of the fields' spatial 
layout. The lattice-like structure of the latter is captured only by those cells, $\{g_1,g_2,\ldots,g_k\}_
\gamma\equiv \varg_{\gamma}$, whose grids were percolated by $\gamma$ and which have therefore produced 
representations, $\mathfrak{G}_{g_1}(\gamma), \mathfrak{G}_{g_2}(\gamma), \ldots, \mathfrak{G}_{g_n}(\gamma)$, 
of $\gamma$ in their respective spiking lattices, $\mathcal{V}_{\mathcal{E}}^{g_1},\mathcal{V}_{\mathcal{E}}^
{g_2}, \ldots,\mathcal{V}_{\mathcal{E}}^{g_k}$. The next path segment, $\gamma'$, is represented by another 
percolated group $\varg_{\gamma'}$ that overlaps with $\varg_{\gamma}$, etc. The resulting series of overlapping
percolated assemblies form a grid cell firing trace $$\digamma_{g}\equiv(\varg_{\gamma},\varg_{\gamma'},\ldots),
$$ that persistently drive hippocampal activity and allow representing longer, composite paths $[\gamma+\gamma'
+\ldots]$. Note that, from the point of view of grid cells' operability, the segments $\gamma,\gamma',\ldots$ 
may overlap and do not necessarily have to extend  across the environment---these assumptions were made above 
for ease of presentation.

A compact bump of persistent head direction activity can then produce congruous deformations of the percolated
path (\ref{shift}) in each contributing lattice, thus generating a compact continuous attractor activity in the
hippocampal network \cite{Romani}. This mechanism allows learning and replaying not only the actual percolating
paths, but also their deformations, thus establishing qualitative equivalences between discretized trajectories
over spiking lattices, facilitating spatial learning, enabling path integration and defining a global spatial 
metric of the encoded environment \cite{Sanders}.

According to the model, the grid cells' percolation onset is modulated by the shape of the navigated arena,
but it is controlled primarily by several coupled physiological parameters---firing rates, field sizes, lattice
spacings, rats' moves and so forth \cite{Watanabe}. Additional restrictions may be required for proper coupling
between different cell types, e.g., place field sizes should allow separating grid fields from each other, for 
encoding distinct vertexes of the spiking lattice $\mathcal{V}_{\mathcal{E}}$. The full set of conditions 
defines a \textit{percolation domain} $\mathcal{P}$ in the parameter space, analogous to the learning region
$\mathcal{L}$ of parameters required for constructing topologically correct cognitive map from place cell 
activity \cite{PLoS,Rev}. An implication of the model is that the experimentally observed spiking 
characteristics should fall into $\mathcal{P}$ and allow producing percolating paths in the amounts required 
for spatial information processing. Certain values can be localized with higher specificity, e.g., the model
predicts that the lattice parameter $\xi_g$ should be attuned to the experimentally observed magnitude $\xi_g^
{\ast}\approx 2/3$, and points at the correct firing rate $A_g\approx 20-25$ Hz in smaller environments 
\cite{Kropff,Hafting,Bush2,MoserM}. Furthermore, the results point out that changes in one parameter may cause
compensatory responses in others, e.g., the network may lower firing rates as $\xi_g$ grows, while producing 
longer percolating paths at a given lattice scale $a_g$ may require growing $A_g$ or using larger fields, i.e.,
shifting the grid cell population activity along the ventro-dorsal axis of MEC. 

\section{Appendix}
\label{sec:met}

\textbf{Simulated trajectories} were obtained by reshaping the recorded rat paths and embedding them into
simulated environments---triangular enclosures of sizes $L=6$ m, $L=12$ m, $L=20$ m and $L=60$ m 
(Fig.~\ref{fig:perc1}). The starting position was selected at the boundary of the enclosure randomly, with
the velocity directed inward. The trajectory was then generated by time-integrating an experimentally recorded 
speed series and directing the velocity vector from one wall to another, with random instantaneous deflections
distributed over an angular domain $[-\alpha,\alpha]$. The parameter $\alpha$ effectively controls the shape 
of the trajectory: small $\alpha$s straighten the paths and larger $\alpha$s allow more ``swirling'' curves. 

\begin{figure}
	\centering
	\includegraphics[scale=0.75]{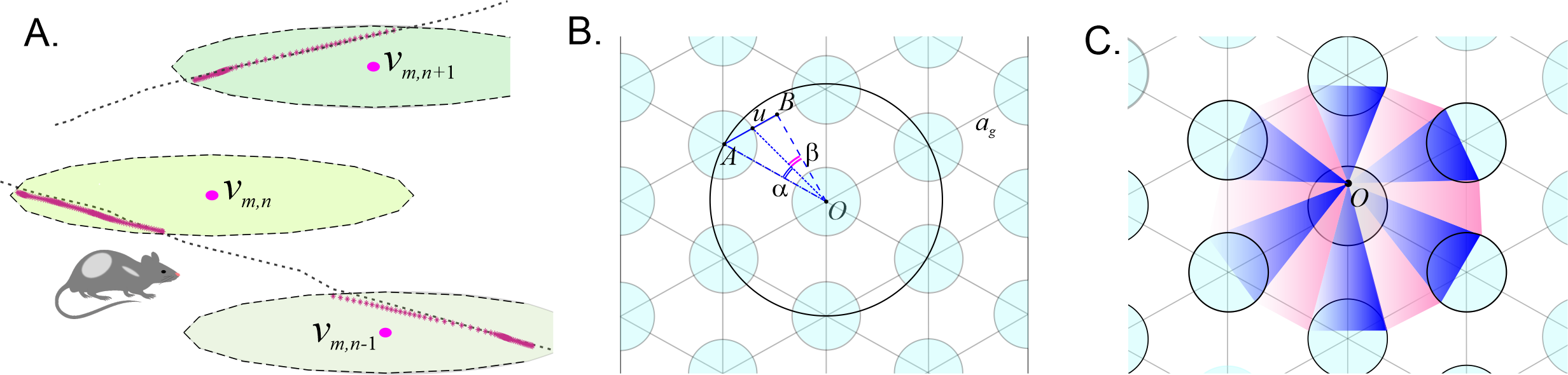}
	\caption{{\footnotesize\textbf{Grid cells}. 
			\textbf{A}. A segment of the rat's trajectory can be approximated by a chord cutting through the
			grid fields. Right panel shows a chord $AB$ of length $l$ passing at the distance $l_{\perp}$ from
			the firing filed center, $v$.
			\textbf{B}. The vertex centered at $A$ may open if the rat moves within the angular domain $\alpha$
			(shaded	blue); if the rat is directed within the domain $\beta$ (pink shade), then the trajectory
			escapes. 
			\textbf{C}. The geometry of the escape changes as the starting point $O$ shifts, leading to small
			corrections to the probability estimate, proportional to the square of distance between $O$ and the
			field center.
	}}
	\label{fig:lgeom}
\end{figure}

\textbf{Site opening probability}. The Poisson firing rate of a grid cell $g$ is a function of the rat's 
position $\vec{r}=(x,y)$
\begin{equation*}
	\lambda_g(\vec{r}) =\sum_{i}A_g e^{-\frac{|\vec{r}-\vec{r}_{i}^{\,g}|^{2}}{2\sigma^{2}_g}},
\end{equation*}
where $A_g$ is the firing amplitude and $\sigma_g$ defines the size of the firing field $\upsilon_{i}^g$
centered at the point $\vec{r}_{i}^{\,g}$. A path segment crossing through $\upsilon_{i}^g$ can be
approximated by a chord of length $l$, parameterized by the variable $u$ and positioned at the distance 
$l_{\perp}$ from the center (Fig.~\ref{fig:lgeom}A,B). The mean integrated rate of the cell $g$ is then
\begin{equation*}
	\bar{\lambda}_g=\int_{\bar{t}_g}\lambda_g dt \approx A_g\int_{AB}e^{-\frac{u^2+l_{\perp}^2}
		{2\sigma_g^2}}\frac{du}{\bar{s}}
	=\frac{A_g}{\bar{s}}e^{-\frac{l_{\perp}^2}{2\sigma_g^2}}\int_{-l/2}^{l/2}e^{-\frac{u^2}{2\sigma_g^2}}du.
	\label{est}
\end{equation*}
Using $D_g\approx2 \pi\sigma_g$ for the firing field diameter, and the relationship $l_{\perp}^2=D_g^2/4-l^2/4$
yield
\begin{equation*}
	\bar{\lambda}_g=\frac{A_g D_g}{\sqrt{2\pi}\bar{s}}e^{-\frac{\pi^2}{2}}e^{\frac{\pi^2l^2}{2D_g^2}}
	\e\left(\frac{\pi l}{\sqrt{2}D_g}\right).
\end{equation*}
From geometric probability theory, the average chord has length
\begin{equation*}
	\bar{l}_g=\pi D_g/4,
\end{equation*}
and hence passes at a distance $\bar{l}_{\perp}\approx 0.31 D_g$ from the field center \cite{Kellerer1}, which
allows writing 
\begin{equation*}
	\bar{\lambda}_g
	=\frac{\bar{l}_g}{\bar{s}}A_g \left(\frac{2}{\pi}\right)^{3/2}e^{-\frac{\pi^2}{2}}e^{\alpha^2\frac{l^2}
		{\bar{l}_g^2}}\e\left(\alpha\frac{l}{\bar{l}_g}\right),
\end{equation*}
where $\alpha=\pi^2/(4\sqrt{2})\approx\sqrt{3}$. The latter equation implies simply that the mean integrated 
rate is proportional to the mean time spent to run through the field, $\bar{t}_g=\bar{l}_g/\bar{s}$. The 
proportionality coefficient between $\bar{\lambda}_g$ and $\bar{t}_g$ can be interpreted as the characteristic
rate during that run,

\begin{equation*}
	\bar{A}_g=A_g \left(\frac{2}{\pi}\right)^{3/2}e^{-\frac{\pi^2}{2}}e^{\alpha^2\frac{l^2}{\bar{l}_g^2}}
	\e\left(\alpha\frac{l}{\bar{l}_g}\right).
\end{equation*}
During an average run, i.e., for $l=\bar{l}_g$, 
\begin{equation*}
	\bar{A}_g\approx 0.0755 A_g,
\end{equation*}
which is equivalent to (\ref{lambdabar}). For example, if the maximal rate is $A_g=25$ Hz, then $\bar{A}_g
\approx 1.8$ Hz (similar values reported in \cite{BrunG}). If the mean speed is $\bar{s}=10$ cm/sec and the
mean filed size is $D_g=40$ cm, then $\bar{t}_g=\bar{l}_g/\bar{s}\approx3$ sec and the net rate is 
$\bar{\lambda}_g\approx\bar{A}_g\bar{t}_g\approx 5.6,$ i.e., the cell spikes with probability $\bar{p}_v^g
\approx 99.6\%$). For $A_g=10$ Hz, vertexes open with probability $\bar{p}_v^g\approx 50\%$.

\textbf{Bond percolation probability}. Consider the case when the rat moves from the center $v$ of a firing
field, outwards along a straight path. The probability $p_b$ of reaching one of the neighboring fields is
defined by the ratio of that field's angular size, as viewed from $v$, and the angular size of the gap between
the firing fields (Fig.~\ref{fig:lgeom}B). Due to symmetries, it is sufficient to consider the domain bounded
by the angle $\angle(AOB)$ and the angles $\alpha\equiv\angle(AOu)$ and $\beta\equiv\angle(uOB)$, $\alpha+\beta
=\frac{\pi}{6}$, which define the probability as
\begin{equation}
	p_b=\frac{\alpha}{\alpha+\beta}=\frac{6\alpha}{\pi}=1-\frac{6\beta}{\pi}.
	\label{lgeom}
\end{equation} 
From the lattice's geometry, $|uB|=(a_g-2R_g)/2$ and $|OB|=\sqrt{3}a_g/2$. From the triangle $AOu$, the distance
$|Ou|$ is $|Ou|^2=R_g^2+a_g^2-R_ga_g$, and from the triangle $uOB$ one has $$|uB|^2=|Ou|^2+\frac{3}{4}a_g^2-|Ou|
a_g\sqrt{3}\cos\beta,$$ which yields
\begin{equation*}
	\cos\beta=\frac{\sqrt{3}a_g}{2\sqrt{R_g^2+a_g^2-R_ga_g}}=\frac{\sqrt{3}}{\sqrt{(\xi_g-1)^2+3}}.
\end{equation*}
For small lattice parameter, $\xi_g\to 0$ (vanishing grid field size), $\beta\to\pi/6$, which eliminates the 
edge opening probability, $p_b(\pi/6) = 0$. Conversely, as the firing field size approaches the gap size, $\xi
_g\to 1$, then the gap vanishes, $\beta\to 0$, which leads to the link opening, $p_b(0)= 1$. The physiological
value $\xi_g^{\ast}\approx 2/3$ produces $\beta^{\ast}\approx0.19$, which corresponds to an overcritical 
probability, $p_b(\beta^{\ast})\approx0.637$. 

If the move starts with an offset $\Delta r$ from the center of the firing field, $r=r_O+\Delta r$, then the
escape probability (\ref{lgeom}) will be an analytical function of $\Delta r/D_g\leq 1$. The zeroth-order term
in the corresponding $(\Delta r/D_g)$-expansion is the mean probability given by (\ref{lgeom}). The first order
term will vanish due to symmetries and the non-vanishing corrections are therefore quadratic, $$P_b(r)=P_b(r_O)
+\partial^2 P_b(\Delta r/D_g)^2/2,$$ which justifies using (\ref{lgeom}) for practical estimates.

\vspace{17pt}
\textbf{Acknowledgments}. The work was supported by the NSF grant 1901338.

\end{document}